\documentclass[copyright,creativecommons,noderivs]{eptcs}

\usepackage{amsmath}
\usepackage{amssymb}
\usepackage{amsbsy}
\usepackage{alltt}
\usepackage{mathpartir}

\newcommand{\adhoc}{\textit{ad hoc}}
\newcommand{\ie}{\textit{i.e.}}

\newcommand{\eg}{\textit{e.g.}}
\newcommand{\konst}[1]{\ensuremath{\mbox{\sf{#1}}}}

\begin{document}

\title{A Step-Indexing Approach to Partial Functions}
\author{David Greve 
\institute{Advanced Technology Center, Rockwell Collins}
\and  Konrad Slind
\institute{Advanced Technology Center, Rockwell Collins}
}

\def\titlerunning{A Step-Indexing Approach to Partial Functions}
\def\authorrunning{D. Greve \& K. Slind}

\maketitle

\begin{abstract}

We describe an ACL2 package for defining partial recursive functions
that also supports efficient execution.  While packages for defining
partial recursive functions already exist for other theorem provers,
they often require inductive definitions or recursion operators which
are not available in ACL2 and they provide little, if any, support for
executing the resulting definitions.  We use step-indexing as the
underlying implementation technology, enabling the definitions to be
carried out in first order logic.  We also show how recent
enhancements to ACL2's guard feature can be used to enable the
efficient execution of partial recursive functions.

\end{abstract}

\section{Introduction}

The provision of support for defining and reasoning about recursive
functions has been an ongoing theme in interactive proof assistants
since at least the time of the original LCF and Boyer-Moore
systems. In most of the current interactive proof systems, one can
expect to be able to introduce a function defined by recursion(s) of
practically any form, and to thereupon be supplied with appropriate
tools for reasoning about the function, \eg, the specified recursion
equations and an induction theorem customized to the recursion
pattern.

Of particular interest in recent work is the ability to define
\emph{partial} functions. Supporting partiality is of course important
in modeling systems of any kind. Partiality also proves to be a
valuable tool when dealing with the well-known difficulties posed by
nested recursion, since partiality support often allows
straightforward proofs of termination \cite{krauss:phd}. Finally,
support for partiality means that the two tasks of (a) defining a
function's behavior and (b) showing that the function terminates, can
be completely separated by first defining the function as a partial
function and, when convenient, showing that the function terminates.

In a logic of total functions, such as ACL2 or HOL, partiality can be
modeled by defining a separate \emph{domain predicate}, which is used
to constrain the introduced function. Given arguments which happen to
be in its domain, the partial function can be unfolded to yield the
corresponding element of its range. A correspondingly constrained
induction theorem can also be produced.

In some work in Isabelle/HOL \cite{krauss:phd} the graph of the
partial function and its domain predicate are introduced by inductive
definitions. However, there are other ways of approaching the
definition task. For example, the first author has shown how
techniques from compilers for functional languages (continuations and
defunctionalization) can be adapted to define partial functions in a
first-order setting \cite{greve:assuming}.  In this paper we explore a
new approach that uses \emph{step-indexing}.  This new approach is
very simple and supports efficient evaluation of partial functions, an
important requirement not previously addressed by others.

\paragraph{Step-indexing}

Step-indexing is a technique which adds a counter to objects being
modeled in a logical construction or an execution. It is typically
used to help in reasoning about difficult recursive constructs, since
the counter helps introduce a notion of step of construction (or
computation), and therefore can sometimes allow simple inductions. By
this expedient, one can often avoid heavyweight domain theory (dealing
with limits of approximations) and instead perform much simpler proofs
on individual approximations (roughly speaking). The nomenclature was
introduced in \cite{appel:mcallester} but similar ideas appear
elsewhere, \eg, in the notion of \emph{interpreter admissibility} used
in the semantics of the ACL2 theory mechanism \cite{ACL2:theories} and 
earlier \cite{boyer_moore:wos-schrift}.

\section{Illustrative Example}

In the following, we will describe the process of definition by a
worked example. Our description may seem somewhat \adhoc{} but, in
fact, the derivations are completely schematic and we are just using
the example to give concrete instances of general proofs.

Ackermann's function will be the running example: it is reasonably
familiar, and has nested recursion, which reveals issues that don't
arise with non-nested recursions. Ackermann's function is defined as:
{\small
\begin{verbatim}
   (equal (ack x y)
          (if (= x 0) (1+ y)
            (if (= y 0) (ack (1- x) 1)
              (ack (1- x) (ack x (1- y))))))
\end{verbatim}
}

This recursion terminates for natural number arguments as it is an
example of \emph{iterated primitive recursion}, \ie, one where the
arguments to each recursive call get smaller under the lexicographic
combination of the predecessor relation on $\mathbb{N}$.  In ACL2,
however, we must consider the behavior of the function for all
possible input values, including the negative integers.  Consequently
the above recursion equations describe a partial function and
\emph{specifies} the following theorems:

\begin{itemize}
\item Equational characterization of the domain.


{\small
\begin{verbatim}
   (equal (ack-domain x y)
          (if (= x 0) t
            (if (= y 0) (ack-domain (1- x) 1)
              (and (ack-domain x (1- y))
                   (ack-domain (1- x) (ack x (1- y)))))))
\end{verbatim}
}

\item Constrained recursion equations:

{\small
\begin{verbatim}
   (implies
     (ack-domain x y)
     (equal (ack x y)
            (if (= x 0) (1+ y)
              (if (= y 0) (ack (1- x) 1)
                (ack (1- x) (ack x (1- y)))))))
\end{verbatim}
}


\item Constrained induction theorem:

{\small
\begin{verbatim}
   (and (implies (and (ack-domain x y)
                      (not (= x 0))
                      (not (= y 0))
                      (:p x (+ -1 y)))
                 (:p x y))
        (implies (and (ack-domain x y)
                      (not (= x 0))
                      (not (= y 0))
                      (:p x (+ -1 y))
                      (:p (+ -1 x) (ack x (+ -1 y))))
                 (:p x y))
        (implies (and (ack-domain x y)
                      (not (= x 0))
                      (= y 0)
                      (:p (+ -1 x) 1))
                 (:p x y))
        (implies (and (ack-domain x y)
                      (= x 0))
                 (:p x y)))
\end{verbatim}
}


\end{itemize}


\section{Base formalization}

We will next show how \konst{ack} and \konst{ack-domain} are defined and
discuss the derivation of the specified theorems. We have considered
two approaches:

\begin{description}
\item [Approach K] transforms the input equations using the partiality
  monad \cite{moggi:notions} before defining a step-indexed version of
  the function. In ACL2, the partial operation is transformed to
  manipulate a pair consisting of an error flag and a value. The
  intended function and its domain are simple definitions in terms of
  the step-indexed version.

\item [Approach G] does not transform the input. Instead, it directly 
   defines the step-indexed function and also the step-indexed domain.
\end{description}

This paper will focus on approach G as it is fully implemented in
ACL2.  The method directly defines \konst{iack}, the step-indexed
version of \konst{ack}, with very little transformation.

{\small
\begin{verbatim}
   (defun iack (d x y)
     (if (zp d) (+ y 1)
       (if (= x 0) (1+ y)
         (if (= y 0) (iack (1- d) (1- x) 1)
            (iack (1- d) (1- x) (iack x (1- d) (1- y)))))))
\end{verbatim}
}


The only difference from the original equations for \konst{ack},
besides the addition of the index $d$ which decrements at each
recursive call, is a new base case which deals with the case when the
index drops to zero. In that case, a default result is returned which
the user may provide.  If the user does not provide a default value,
the mechanization picks a value from among the original base cases
($y+1$ in this example).

\paragraph {Indexed domain}

The logical domain predicate \konst{Lack-dom}, to be introduced later, is
defined in terms of a step-indexed version \konst{iack-dom}, which is,
again, defined primitive recursively over an index that decrements at
each recursive call.

{\small
\begin{verbatim}
   (defun iack-dom (d x y)
     (if (zp d) (= x 0)
       (if (= x 0) t
         (if (= y 0) (iack-dom (1- d) (1- x) 1)
            (and (iack-dom (1- d) x (1- y))
                 (iack-dom (1- d) (1- x) (iack (1- d) x (1- y))))))))
\end{verbatim}
}


Note, however, that the definition of \konst{iack-dom} is not a completely
direct adaptation of the original equations, since inner calls in
nested recursions are lifted out separately as arguments to
\konst{iack-dom}.  Again, we have to supply a value when the index drops
to zero.  A disjunction of all of the tests that drive control flow
into a base case is used ($x=0$ in our example).

\paragraph {Measure function}

A crucial part of the development is a \emph{measure} function,
\konst{ack-measure}, having the property that it yields the least
depth of recursion needed to obtain a result for the given inputs,
when such a depth exists.  We proceed in two steps. First, we use the
{\small\verb+defchoose+} facility to introduce a function
\konst{ack-depth}, which yields a depth of recursion sufficient to obtain
a result if such a depth exists.

{\small
\begin{verbatim}
  (defchoose ack-depth (d)
    (iack-dom d x y))
\end{verbatim}
}


\noindent While this step of the formalization is not constructive and
not computable we will derive computable consequences.

In order to formulate the desired induction theorem for \konst{ack},
we will need to have the \emph{least} depth of recursion. The least
depth can be computed once it is known that there is a depth at which
recursion terminates; thus we construct a recursive function
\konst{iack-min-index} which returns the smallest depth at which
\konst{iack-dom} holds.

{\small
\begin{verbatim}
   (defun iack-min-index (d x y)
     (if (zp d) 0
       (if (not (iack-dom d x y)) 0
         (if (not (iack-dom (1- d) x y)) d
           (iack-min-index (1- d) x y)))))
\end{verbatim}
}

With that in hand, we define

{\small
\begin{verbatim}
   (defun ack-measure (x y)
     (iack-min-index (ack-depth x y) x y))
\end{verbatim}
}


and prove that \konst{ack-measure} returns the least index, when there
is an index at which \konst{iack} terminates. We then define the
logical definition for \konst{ack}---\konst{Lack}---and a logical
definition for its domain---\konst{Lack-dom}.

{\small
\begin{verbatim}
  (defun Lack (x y) (iack (ack-measure x y) x y))
  
  (defun Lack-dom (x y) (iack-dom (ack-measure x y) x y))
\end{verbatim}
}


\paragraph{Basic properties}

Later proofs require a small collection of properties about the
definedness of \konst{iack}, namely that it is deterministic and
stable, and \konst{ack-measure} is canonical.

{\small
\begin{verbatim}
   (defthm iack-deterministic
     (implies (and (iack-dom d1 x y) (iack-dom d2 x y))
              (equal (iack d1 x y)
                     (iack d2 x y))))

   (defthm iack-stable
     (implies (and (iack-dom d1 x y) (< (nfix d1) (nfix d2)))
              (iack-dom d2 x y)))

   (defthm iack-measure-canonical
     (implies (iack-dom d x y)
              (equal (iack d x y)
                     (iack (ack-measure x y) x y))))
\end{verbatim}
}


\noindent These are straightforward to prove.

In ACL2, the least-depth property possessed by \konst{ack-measure} is
not as useful to the proof automation as a recursive characterization,
especially in the proofs of the recursive presentations of \konst{Lack-dom}
and \konst{Lack}. Here is the recursion equation for \konst{ack-measure}:

{\small
\begin{verbatim}
   (equal (ack-measure x y)
          (if (not (Lack-dom x y)) 0
            (if (= x 0) 0
              (if (= y 0) (1+ (ack-measure (1- x) 1))
                (1+ (max (ack-measure x (1- y))
                         (ack-measure (1- x) (Lack x (1- y)))))))))
\end{verbatim}
}


\noindent Thus, the depth of a recursive call is one less than that of
the originating call. For multiple recursions, the maximum of the
depths of the recursions is one less than the depth of the originating
call.  As a result, the measure decreases along each recursive call.
Using this equation for \konst{ack-measure}, ACL2 is able to prove the
following theorems about \konst{Lack-dom} and \konst{Lack}:

{\small
\begin{verbatim}
   (equal (Lack-dom x y)
          (if (= x 0) t
            (if (= y 0) (Lack-dom (1- x) 1)
              (and (Lack-dom x (1- y))
                   (Lack-dom (1- x) (Lack x (1- y)))))))
\end{verbatim}
}

{\small
\begin{verbatim}
   (equal (Lack x y)
          (if (not (Lack-dom x y)) (1+ y)
            (if (= x 0) (1+ y)
              (if (= y 0) (Lack (1- x) 1)
                (Lack (1- x) (Lack x (1- y))))))
\end{verbatim}
}


With the characterizations of the domain and the measure, we are now
in a position to introduce a logical induction scheme for
\konst{Lack}.  The induction scheme is a variation of the body of
\konst{Lack-dom}, extended with a guard on the domain, \konst{Lack-dom}, and
justified by \konst{ack-measure}.

{\small
\begin{verbatim}
   (defun Lack-induction (x y)
     (declare (xargs :measure (ack-measure x y)))
     (if (not (Lack-domain x y)) nil
       (if (= x 0) t
         (if (= y 0) (Lack-induction (1- x) 1)
           (and (Lack-induction x (1- y))
                (Lack-induction (1- x) (Lack x (1- y))))))))
\end{verbatim}
}


\section{Executable versions}

We now have a useful logical theory for \konst{Lack}: a defining
theorem, a domain predicate, a measure and an induction scheme.
Constructing efficient executables within this logical theory,
however, is not simple.  The defining theorem for \konst{Lack}
includes a call of \konst{Lack-dom} and the defining theorem for
\konst{Lack-dom} includes a call of \konst{Lack}.  Thus, \konst{Lack}
and \konst{Lack-dom} are mutually recursive.  However, naively
checking membership in the domain for every argument in an execution
would be unnecessarily expensive.  To address this we define a
mutually recursive set of functions, \konst{mack} (for mutually
recursive \konst{ack}) and \konst{ack-domain}, and attach executable
bodies to them using appropriate guards and \konst{MBE}. The logical
definitions of these functions are quite benign; it is in the
executable definitions and the guards that things get interesting.

{\small
\begin{verbatim}
   (mutual-recursion

     (defun mack (x y)
       (declare (xargs :guard (ack-domain x y)))
       (mbe :logic (Lack x y)
            :exec  (if (= x 0) (1+ y)
                     (if (= y 0) (mack (1- x) 1)
                       (mack (1- x) (mack x (1- y)))))))

     (defun ack-domain (x y)
       (mbe :logic (Lack-dom x y)
            :exec  (if (= x 0) t
                     (if (= y 0) (ack-domain (1- x) 1)
                       (and (ack-domain x (1- y))
                            (ack-domain (1- x) (Lack x (1- y))))))))

     )
\end{verbatim}
}



The executable body of \konst{ack-domain} still calls \konst{mack},
but note that the executable body of \konst{mack} does not call
\konst{ack-domain}.  A check on the domain is, however, necessary to
complete the requisite proof that the logical body, \konst{Lack}, is
the same as the executable body.  This necessary check on the domain
is included as a call to \konst{ack-domain} in the guard of
\konst{mack}\footnote{In addition to any user provided guards}.  The
ability to use other functions from within a mutually recursive clique
as guards was added to ACL2 in version 3.6.  The beauty of using
\konst{ack-domain} as a guard is that it need only be satisfied once
(prior to calling \konst{mack}) rather than once every iteration of
\konst{mack}, as would have been the case had we used the defining
theorem of \konst{Lack} above.

While a single call to \konst{ack-domain} is much better than many
calls, any such call can result in substantial execution overhead for
a given call of \konst{ack}.  For reflexive functions such as
\konst{ack}, the cost of evaluating the domain function may actually
be exponentially more expensive than the cost of evaluating the
function itself.  To minimize this overhead, we refine our executable
model even further by introducing another indexed version of
\konst{ack}, called \konst{comp-ack} for \emph{computational} \konst{ack}.
The only difference between \konst{comp-ack} and \konst{iack} is in
the case when the bound $d$ is exhausted: \konst{iack} simply returns
the default value, while \konst{comp-ack} checks \konst{ack-domain}
and, if false, returns the default value, but if true, continues
execution by calling \konst{mack}.  Note that this domain check
satisfies the guards of \konst{mack}.

{\small
\begin{verbatim}
   (defun comp-ack (d x y)
     (if (zp d) (if (ack-domain x y) (mack x y) (+ y 1))
       (if (= x 0) (1+ y)
         (if (= y 0) (comp-ack (1- d) (1- x) 1)
            (comp-ack (1- d) (1- x) (comp-ack x (1- d) (1- y)))))))
\end{verbatim}
}


In this function, the wasted computation of the domain check is
deferred: the function runs nearly as fast as possible, with only the
addition of the index decrement and check, until the index bound is
exhausted.  Only then does it perform a potentially expensive domain
check.  If the arguments are in the domain, execution completes as
quickly as possible without any further domain checks or index
counters.  Now we pick some large constant number \konst{BIG}---in our
current implementation, the largest number fitting into a machine
integer---and finally make the ultimate definition of \konst{ack}.

{\small
\begin{verbatim}
   (defun ack (x y)
     (comp-ack (BIG) x y))
\end{verbatim}
}


This function allows us to compute values of the partial function
\konst{ack} by invoking \konst{comp-ack}. The only slowdown until the
bound is reached is the constant cost of decrementing the index at
each recursive call. \konst{BIG} is large enough, especially
in the era of 64-bit machine integers, that most applications that
will terminate should terminate long before the index is exceeded.

The following characterization of \konst{ack} is then provable:

{\small
\begin{verbatim}
  (equal (ack x y)
         (if (not (ack-domain x y)) (comp-ack (BIG) x y)
           (if (= x 0) (1+ y)
            (if (= y 0) (ack (1- x) 1)
              (ack (1- x) (ack x (1- y)))))))
\end{verbatim}
}


Note that the behavior of \konst{ack} outside of the function domain
is not simply our default value, but has been complicated by our use
of \konst{comp-ack}.  There is certainly a trade-off here between
execution efficiency and simplicity in that case.  The above
characterization of \konst{ack} is what we export, along with updated
versions of \konst{ack-domain} and \konst{ack-measure} expressed in
terms of \konst{ack} and a final induction scheme,
\konst{ack-induction}, also defined with respect to \konst{ack} and
justified by \konst{ack-measure}. This gives us the desired
combination of logical reasoning power plus fast computations via
\konst{comp-ack}.

\section{Implementation}

A mechanized implementation of these ideas has been codified in ACL2
in a macro called \konst{def::ung}.  While our previous discussion
illustrated the general behavior of this macro, the actual behavior
exhibited by \konst{def::ung} for a given invocation may be a subset
of the behaviors we describe above.  Nonetheless, the macro is fully
automated and is designed to behave as a replacement for \konst{defun}
for introducing partial recursive functions.

Just as with \konst{defun}, \konst{def::ung} constructs guards from
Common Lisp declarations and the \konst{xargs} \konst{:guard} keyword.
It also deduces guard conditions from the \konst{xargs}
\konst{:signature} keyword.  This feature, inherited from
\konst{def::un} (from coi/util/defun in the ACL2 books), provides a
convenient, Common Lisp declaration-inspired language for specifying a
function's logical signature.  Guard proofs can be controlled using
\konst{:guard-hints} and the signature proofs can be controlled using
\konst{:signature-hints}.  Such low-level control is, however,
discouraged.  A better approach is to admit the definition in a theory
conducive to automated proofs of these conjectures.  Guard proofs can
also be delayed using \konst{:verify-guards} nil.  Note that guard
verification of the admitted function may require guard verification
of a number of supporting functions as well.

If no guard information is provided by the user in the form of
declarations or the \konst{:guard} or \konst{:signature} keywords,
\konst{def::ung} will produce an executable from the logical
definition using \konst{ec-call} to suppress any residual guard
conditions.  Such default execution behavior roughly mimic that of
\konst{defun}.  Note that when no guard information is provided to
\konst{def::ung}, an indexed executable is not generated under the
rationale that efficient execution is not a priority in that case.

In addition to the \konst{:signature} and \konst{:signature-hints}
keywords, the \konst{def::ung} macro accepts a number of other
non-standard \konst{xargs} keywords that give the user additional
control over its behavior.

\begin{description}
\item{\konst{:default-value} \emph{expr}}

The \konst{:default-value} keyword allows the user to provide an
expression to compute the default value to be returned by the function
when its arguments are outside of the function domain.  This
expression may be computed from the function arguments.  When
\konst{:default-value} is not specified, \konst{def::ung} chooses a
default value from among the function's base cases.

\item{\konst{:non-executable} [nil]/t}

\konst{Def::ung} will, by default, attempt to produce an executable
function definition.  When \konst{:non-executable} is t,
\konst{def::ung} will provide only the logical theory for the function
with absolutely no support for execution.  Nonetheless, if a
\konst{:signature} is provided, an appropriate type theorem will still
be generated.

\item{\konst{:indexed-execution} [t]/nil}

By default, when guard information is provided, the executable
function defined by \konst{def::ung} is indexed to optimize
performance.  The indexed function is, however, somewhat more
difficult to reason about outside of the function's domain.  If this is
an issue, the user may specify \konst{:indexed-execution} nil to
suppress the generation and use of an indexed execution function.

\item{\konst{:wrapper-macro} \emph{name}}

When \konst{:indexed-execution} is nil, the guard of the resulting
executable requires that the function arguments be in the domain of
the function.  When a name is provided via the \konst{:wrapper-macro}
keyword, \konst{def::ung} will generate a wrapper macro that tests
whether the arguments are in the domain prior to calling the function.
If the arguments are not in the domain, the default value is returned.
This wrapper macro can then be invoked in place of the function when
the domain guard may not be satisfied.
\end{description}

Following is an illustration of how we might define our \konst{ack}
example using \konst{def::ung}.  Because we provide guard information,
\konst{def::ung} will generate an indexed executable.  The default
value we provide here, 0, is used to define the behavior of the
function on arguments outside of its domain, as in (ack -1 0).

{\small
\begin{verbatim}
   (def::ung ack (x y)
     (declare (xargs :signature ((natp natp) natp) 
                     :default-value 0))
     (if (= x 0) (1+ y)
       (if (= y 0) (ack (1- x) 1)
         (ack (1- x) (ack x (1- y))))))
\end{verbatim}
}

\noindent We can verify the defining equation of \konst{ack}:

{\small
\begin{verbatim}
   (defthm check-ack-definition
     (equal (ack x y)
            (if (not (ack-domain x y)) (ack-compute (defung::big-depth-fn) x y)
              (if (= x 0) (1+ y)
                (if (= y 0) (ack (1- x) 1)
                  (ack (1- x) (ack x (1- y)))))))
     :hints (("Goal" :in-theory (disable (:rewrite defung::generalize-big-depth-fn))))
     :rule-classes nil)
\end{verbatim}
}

\noindent We can also execute \konst{ack} on some concrete values:

{\small
\begin{verbatim}
ACL2 !>(time$ (ack 3 11))
   ; (EV-REC *RETURN-LAST-ARG3* ...) took 
   ; 1.25 seconds realtime, 1.25 seconds runtime
   ; (1,120 bytes allocated).
   16381
\end{verbatim}
}

\noindent Here we run an equivalent program mode definition for
comparison:

{\small
\begin{verbatim}
   ACL2 !>(defun ack0 (x y)
     (declare (xargs :mode :program))
     (if (= x 0) (1+ y)
       (if (= y 0) (ack0 (1- x) 1)
         (ack0 (1- x) (ack0 x (1- y))))))

   Summary
   Form:  ( DEFUN ACK0 ...)
   Rules: NIL
   Time:  0.00 seconds (prove: 0.00, print: 0.00, other: 0.00)
    ACK0
   ACL2 !>(time$ (ack0 3 11))
   ; (EV-REC *RETURN-LAST-ARG3* ...) took 
   ; 0.61 seconds realtime, 0.61 seconds runtime
   ; (1,120 bytes allocated).
   16381
\end{verbatim}
}

It is interesting to observe the impact that the domain check has on
computational performance.  Here we define a version of \konst{ack}
without indexed computation.  When evaluating this function in the top
level loop, ACL2 first checks whether the inputs are in the domain of
the function by executing \konst{ack2-domain}.

{\small
\begin{verbatim}
  (def::ung ack2 (x y)
    (declare (xargs :signature ((natp natp) natp)
                    :indexed-execution nil
                    :wrapper-macro ack2-exec))
    (if (= x 0) (1+ y)
      (if (= y 0) (ack2 (1- x) 1)
        (ack2 (1- x) (ack2 x (1- y))))))
\end{verbatim}
}

Now compare the difference in time between evaluating \konst{ack2} and
\konst{ack2-domain} below.  Note that the \konst{ack2-domain}
computation time dominates the \konst{ack2} computation time and that
\konst{ack}, which executes without a domain check, essentially makes
up the difference.  With reflexive functions such as \konst{ack2}, it
is possible for the domain computation time to grow exponentially
faster than the function computation time.

{\small
\begin{verbatim}
   ACL2 !>(time$ (ack2 3 8))
   ; (EV-REC *RETURN-LAST-ARG3* ...) took 
   ; 15.34 seconds realtime, 15.34 seconds runtime
   ; (1,120 bytes allocated).
   2045
   ACL2 !>(time$ (ack2-domain 3 8))
   ; (EV-REC *RETURN-LAST-ARG3* ...) took 
   ; 15.32 seconds realtime, 15.31 seconds runtime
   ; (1,120 bytes allocated).
   T
   ACL2 !>(time$ (ack 3 8))
   ; (EV-REC *RETURN-LAST-ARG3* ...) took 
   ; 0.02 seconds realtime, 0.02 seconds runtime
   ; (1,120 bytes allocated).
   2045
\end{verbatim}
}

The support for partiality provided by \konst{def::ung} means that the
two tasks of (a) defining a function's behavior and (b) showing that
the function terminates, can be completely separated by first defining
the function as a partial function and, when convenient, showing that
the function terminates.  The \konst{def::total} macro, also exported
by the \konst{def::ung} book, provides support for proving termination
of functions admitted using \konst{def::ung}.  The macro allows the
user to specify an \konst{xargs} \konst{:measure} and, as appropriate,
a \konst{:well-founded-relation} that justifies termination.  In
addition, the body of the macro may contain a predicate that
articulates the condition under which the function is total.  This
condition may simply be t.  If multiple termination proofs (presumably
under different conditions) are desired, the user may specify
different names for the different proofs using the \konst{xargs}
keyword \konst{:totality-theorem} \emph{name}.

We can prove the totality of our example, \konst{ack}, when the
function guards are satisfied, that is: when the inputs are natural
numbers.

{\small
\begin{verbatim}
   (def::total ack (x y)
     (declare (xargs :measure (llist x y)
                     :well-founded-relation l<
                     :totality-theorem natp-ack-terminates))
     (and (natp x) (natp y)))
\end{verbatim}
}

The \konst{def::ung} macro is being used within Rockwell Collins to
develop tools for reasoning about software systems, especially systems
written in the \konst{C} programming language.  While the correctness
of such tools is of interest to us, their termination is not.
Additionally, because these tools are being applied to actual
\konst{C} programs, it is important that they execute quickly.  The
\konst{def::ung} macro addresses both of these issues, allowing us to
reason formally about our tools within the logic and allowing them to
execute quickly on concrete input.

\section{Related work}

The second author's PhD \cite{slind:phd} developed a recursion
operator-based approach to defining total recursive functions. One
theme in the work was attempting to separate the definition of a
function from reasoning about its termination, a long-standing problem
with nested recursions.  Although successful for total functions, the
approach failed to capture any notion of partiality or explicit domain
of a function and hence nested recursive functions were painful to
formalize and reason about, and partial functions were not dealt with
at all.

In his dissertation work, Krauss \cite{krauss:phd} took a different
tack.  Instead of instantiating a pre-proved recursion theorem, a
construction is done fresh for each function submitted. The approach
uses inductive definitions to construct the graph and the domain of
the function. The specified function itself is obtained once the graph
is (automatically) shown to be functional, and then the constrained
induction theorem and recursion equations are derived. The package
also comes with support for automated termination proofs.

A nice overview---as of 2006--of support in Type Theory
implementations for recursive definitions is given in
\cite{barthe:recdef}. The technical contribution in the paper is based
on inductively defining the graph, similar to (and contemporaneous
with) Krauss' approach, although it did not deal with nested recursion.

ACL2 for a long time only supported total functions, but Manolios
and Moore \cite{manolios_moore:pfun} discovered that tail-recursions
are consistent to admit into ACL2. Based on that work, the first
author of the present paper created an ACL2 macro that maps arbitrary
recursive specifications into CPS (continuation-passing style) and
then transforms the CPS result down to first order, obtaining a
tail-recursive model of the original function, from which the desired
equations and proof principles can be derived \cite{greve:assuming}.
Similar to Krauss' work, a separate domain predicate is defined, thus
separating the definition of a function and its termination proof.
The implementation was, however, complex and performed somewhat
inefficiently.

The general approach taken in TFL \cite{slind:phd}, namely to
instantiate a pre-proved recursion theorem, compresses much of the
model-building work into one theorem that can be instantiated and
manipulated to deliver the desired result. It remains to be seen
whether a similarly useful recursion theorem can be generated for
partial functions, whereby the domain of the function is explicitly
included. This could be a potential application of encapsulate and
functional instantiation in ACL2.

Finally, none of the work we are aware of deals with the issue of
execution of partial functions in theorem provers.

\section{Future Work}

The ability to admit partial functions and to delay proofs of
termination is a useful feature.  The \konst{def::total} macro already
provides support for proving the termination of partial functions
admitted with \konst{def::ung}.  Total functions, however, can be
executed in ACL2 more efficiently than partial functions.  The ability
to add an executable body to an encapsulated function symbol or to
replace an existing executable body has been recently added to ACL2
(via \konst{defattach}).  A useful extension of \konst{def::total}
would be the ability to attach a more efficient (total) executable
body to the original partial function symbol.

Ideally \konst{def::ung} would be a seamless replacement for
\konst{defun}.  While the current package attempts this, certain
aspects of function admission in ACL2 are out of our control.  For
example, while it is possible to assign theorems to the rule class
\konst{:definition}, it is not possible to control the names of runic
designators.  We cannot, for example, change the runic designator used
by \konst{defun} and we cannot assign the runic designator
\konst{(:definition foo)} to a theorem whose name is not \konst{foo}.
Without such ability it is impossible for a user to mimic fully the
behavior of built-in macros such as \konst{defun}.  The clever use of
\konst{add-macro-alias} (as in \konst{defun-inline}), however, might
strengthen the desired illusion.

Single threaded objects and multi-value returns are always an issue
for macros that generate or manipulate function definitions.  They are
particularly bothersome because they do not readily admit generic
solutions, going so far as to infect even such constructs as
encapsulation.  The \konst{def::ung} macro does not currently support
either construct.

Early experiments suggest that the monadic approach (Approach K),
while requiring more extensive surgery to the body of the function
definition, may substantially improve execution speed by executing the
domain computation in parallel with the function computation. The
primary drawback of this approach is that, for optimal efficiency, it
requires the use of multiple-value returns -- a construct not
currently supported in the framework.  Nonetheless, this approach may
ultimately be required in order to provide general support for
single-threaded objects since the domain computation in such functions
will likely involve predicates over an evolving single-threaded state.

Care has been taken to automate, control, and streamline the proof
process behind \konst{def::ung}.  All of the proofs performed by the
macro are schematic, meaning that they follow the same line of
reasoning with every invocation.  Unfortunately, it is often difficult
to keep ACL2 on the script.  For example, ACL2 will always replace
symbols with nil if it knows that the symbol is null.  Unfortunately,
even such simple transformations can break our schematic proofs.  ACL2
also has difficulty manipulating large single-threaded function
definitions and it is nearly impossible to keep ACL2 from performing
beta reduction.  As a result, \konst{def::ung} can be much slower and
more brittle than we would like.  To address such issues, we are
exploring the possibility of using clause processors to isolate and
insulate the proof process from the whims of ACL2.

However, the ultimate solution might be to simply incorporate a
partial function definition capability directly into the ACL2 core.
While the definitions generated by \konst{def::ung} have been verified
to be sound on a variety of examples, it is unclear to the authors how
one might verify the soundness of the approach once and for all.


\section{Conclusions}

We have developed a new ACL2 package for partial function definition,
with particular emphasis on efficient execution of partial functions.
A long-term goal of ours has been to move programming notions as much
as possible into logic, so that our chosen logic environments allow
the full comfort of programming with the usual idioms and techniques
while also providing direct and unfettered use of the theorem prover
to establish properties. An important part of that goal is efficient
execution of formal models while maintaining a strong connection
between the function as an entity being reasoned about and the
function as an entity being executed. The package discussed in this
paper supports programming as an activity that can be done inside a
theorem prover without sacrificing execution speed or burdening the
programmer with onerous termination proofs.

\subsubsection*{Acknowledgments}

It is a pleasure to acknowledge Matt Kaufmann's major contributions to
this research. Many of the ideas presented here have been learned from
discussions with Matt and in extended sessions at the terminal with
him.

\bibliographystyle{eptcs}
\bibliography{partial}

\begin{thebibliography}{1}
\providecommand{\bibitemdeclare}[2]{}
\providecommand{\surnamestart}{}
\providecommand{\surnameend}{}
\providecommand{\urlprefix}{Available at }
\providecommand{\url}[1]{\texttt{#1}}
\providecommand{\href}[2]{\texttt{#2}}
\providecommand{\urlalt}[2]{\href{#1}{#2}}
\providecommand{\doi}[1]{doi:\urlalt{http://dx.doi.org/#1}{#1}}
\providecommand{\bibinfo}[2]{#2}

\bibitemdeclare{article}{appel:mcallester}
\bibitem{appel:mcallester}
\bibinfo{author}{Andrew~W. \surnamestart Appel\surnameend} \&
  \bibinfo{author}{David \surnamestart McAllester\surnameend}
  (\bibinfo{year}{2001}): \emph{\bibinfo{title}{An indexed model of recursive
  types for foundational proof-carrying code}}.
\newblock {\sl \bibinfo{journal}{ACM Trans. Program. Lang. Syst.}}
  \bibinfo{volume}{23}(\bibinfo{number}{5}), pp. \bibinfo{pages}{657--683},
  \doi{10.1145/504709.504712}.

\bibitemdeclare{incollection}{barthe:recdef}
\bibitem{barthe:recdef}
\bibinfo{author}{Gilles \surnamestart Barthe\surnameend},
  \bibinfo{author}{Julien \surnamestart Forest\surnameend},
  \bibinfo{author}{David \surnamestart Pichardie\surnameend} \&
  \bibinfo{author}{Vlad \surnamestart Rusu\surnameend} (\bibinfo{year}{2006}):
  \emph{\bibinfo{title}{Defining and Reasoning About Recursive Functions: A
  Practical Tool for the Coq Proof Assistant}}.
\newblock In \bibinfo{editor}{Masami \surnamestart Hagiya\surnameend} \&
  \bibinfo{editor}{Philip \surnamestart Wadler\surnameend}, editors: {\sl
  \bibinfo{booktitle}{Functional and Logic Programming}}, {\sl
  \bibinfo{series}{Lecture Notes in Computer Science}} \bibinfo{volume}{3945},
  \bibinfo{publisher}{Springer Berlin Heidelberg}, pp.
  \bibinfo{pages}{114--129}, \doi{10.1007/11737414\_9}.

\bibitemdeclare{inbook}{boyer_moore:wos-schrift}
\bibitem{boyer_moore:wos-schrift}
\bibinfo{author}{Robert~S. \surnamestart Boyer\surnameend} \&
  \bibinfo{author}{J~Strother \surnamestart Moore\surnameend}
  (\bibinfo{year}{1997}): \emph{\bibinfo{title}{Automated reasoning and its
  applications}}, chapter \bibinfo{chapter}{Mechanized formal reasoning about
  programs and computing machines}, pp. \bibinfo{pages}{147--176}.
\newblock \bibinfo{publisher}{MIT Press}, \bibinfo{address}{Cambridge, MA,
  USA}.
\newblock \urlprefix\url{http://dl.acm.org/citation.cfm?id=271101.271126}.

\bibitemdeclare{inproceedings}{greve:assuming}
\bibitem{greve:assuming}
\bibinfo{author}{David \surnamestart Greve\surnameend} (\bibinfo{year}{2009}):
  \emph{\bibinfo{title}{Assuming termination}}.
\newblock In: {\sl \bibinfo{booktitle}{Proceedings of the Eighth International
  Workshop on the ACL2 Theorem Prover and its Applications}},
  \bibinfo{series}{ACL2 '09}, \bibinfo{publisher}{ACM}, \bibinfo{address}{New
  York, NY, USA}, pp. \bibinfo{pages}{114--122}, \doi{10.1145/1637837.1637856}.

\bibitemdeclare{article}{ACL2:theories}
\bibitem{ACL2:theories}
\bibinfo{author}{Matt \surnamestart Kaufmann\surnameend} \&
  \bibinfo{author}{J.~Strother \surnamestart Moore\surnameend}
  (\bibinfo{year}{2001}): \emph{\bibinfo{title}{Structured Theory Development
  for a Mechanized Logic}}.
\newblock {\sl \bibinfo{journal}{J. Autom. Reason.}}
  \bibinfo{volume}{26}(\bibinfo{number}{2}), pp. \bibinfo{pages}{161--203},
  \doi{10.1023/A:1026517200045}.

\bibitemdeclare{article}{krauss:phd}
\bibitem{krauss:phd}
\bibinfo{author}{Alexander \surnamestart Krauss\surnameend}
  (\bibinfo{year}{2010}): \emph{\bibinfo{title}{Partial and Nested Recursive
  Function Definitions in Higher-order Logic}}.
\newblock {\sl \bibinfo{journal}{J. Autom. Reason.}}
  \bibinfo{volume}{44}(\bibinfo{number}{4}), pp. \bibinfo{pages}{303--336},
  \doi{10.1007/s10817-009-9157-2}.

\bibitemdeclare{article}{manolios_moore:pfun}
\bibitem{manolios_moore:pfun}
\bibinfo{author}{Panagiotis \surnamestart Manolios\surnameend} \&
  \bibinfo{author}{J.~Strother \surnamestart Moore\surnameend}
  (\bibinfo{year}{2003}): \emph{\bibinfo{title}{Partial Functions in ACL2}}.
\newblock {\sl \bibinfo{journal}{J. Autom. Reason.}}
  \bibinfo{volume}{31}(\bibinfo{number}{2}), pp. \bibinfo{pages}{107--127},
  \doi{10.1023/B:JARS.0000009505.07087.34}.

\bibitemdeclare{article}{moggi:notions}
\bibitem{moggi:notions}
\bibinfo{author}{Eugenio \surnamestart Moggi\surnameend}
  (\bibinfo{year}{1991}): \emph{\bibinfo{title}{Notions of computation and
  monads}}.
\newblock {\sl \bibinfo{journal}{Inf. Comput.}}
  \bibinfo{volume}{93}(\bibinfo{number}{1}), pp. \bibinfo{pages}{55--92},
  \doi{10.1016/0890-5401(91)90052-4}.

\bibitemdeclare{phdthesis}{slind:phd}
\bibitem{slind:phd}
\bibinfo{author}{Konrad \surnamestart Slind\surnameend} (\bibinfo{year}{1999}):
  \emph{\bibinfo{title}{Reasoning about Terminating Functional Programs}}.
\newblock Ph.D. thesis, \bibinfo{school}{Institut f{\"u}r Informatik,
  Technische Universit{\"a}t M{\"u}nchen}.

\end{thebibliography}

\end{document}